# Low Frequency Raman Scattering from Acoustic Phonon Confined in $CdS_{1-x}Se_x$ Nanoparticles in Borosilicate Glass

Sanjeev K. Gupta, Satyaprakash Sahoo[*], Prafulla K. Jha, A. K. Arora[*], Y. M. Azhniuk[#]

Nanotechnology Centre, Department of Physics, Bhavnagar University,
Bhavnagar, 364 002, India.
[*]Light Scattering Section Studies, Material Science Division,
Indira Gandhi Centre for Atomic Research, Kalpakkam, Tamilnadu, 602103, India.
[#]Institute of Electron Physics, Ukr. Nat. Acad Sci., Universytetska St. 21,
Uzhhorod 88000, Ukraine.

## ABSTRACT

Phonon modes found in low frequency Raman scattering from $CdS_{1-x}Se_x$ nanocrystals embedded in borosilicate glass arising from confined acoustic phonons are investigated. In addition to the breathing modes and quadrupolar modes, two additional modes are found in the spectra. In order to assign the new modes, confined acoustic phonon frequencies are calculated using CFM, CSM and the Lamb model. Based on the ratio of the frequencies of the new modes to those of the quadrupolar mode, the new modes are assigned to first overtone of the quadrupolar mode ($l$=2, $n$=1) and $l$=1, $n$=0 torsional mode. The appearance of the forbidden torsional mode is attributed to nonspherical appearance shape of the nanoparticle found from high-resolution TEM.

## I. INTRODUCTION

In the last two decades, the glass-embedded $CdS_{1-x}Se_x$ semiconductor quantum dots have attracted considerable attention due to their significantly different behavior from those of the corresponding bulk crystals [1-7]. They are interesting due to the size-dependent effects related to spatial confinement of charge carrier and their applications as optical processing devices [1-2, 8-12]. The chemical compositions of mixed nanocrystals as well as their size are the main parameters determining the properties of glass-embedded quantum dots. To understand the optical properties of semiconductor nanocrystals, it is absolutely necessary to consider the zero dimensional confinement effects on the electronic states in a system. Raman spectroscopy has emerged as a powerful and non destructive technique for estimating the composition of glass embedded nanocrystal [13-14] and their size [15-16]. Raman scattering probes the vibrational states of the particles such as the confined acoustic, optical phonons and surface phonon modes. It also provides an understanding of electron-phonon interactions. This has led to several investigations of acoustic vibrations of spherical nanoparticles, particularly the semiconductor nanoparticles leading to an understanding of the role of vibrations in the performance of some optical devices (for example, in electronic dephasing due to emission of phonons) [17-18]. The low frequency phonon modes (acoustic phonons) of nanoparticles bear a unique signature of their structural and mechanical properties besides modifying the electronic structure. There exist several studies on low frequency Raman scattering from elastic spherical nanoparticles vibrating with frequencies inversely proportional to their size. Acoustic vibrations of the elastic sphere having frequency very low compared to laser light lead to variations in the dielectric response of the sphere to the light. As the sphere slowly vibrates, its dielectric response leads to Raman or Brillouin scattered light whose frequency is slightly above or below that of the incident laser. To describe these vibrations usually the continuum elastic models have been used. The original 1882 work by Lamb [19] is the basis for these models which describes the vibrations as eigenfrequencies of a homogeneous elastic sphere under stress-free boundary conditions. The frequencies are classified into two categories, spheroidal and torsional; torsional modes being Raman inactive. In most of the studies on the low frequency Raman scattering of $CdS_{1-x}Se_x$, only the spheroidal modes of *l*=0 (symmetrical mode) and for *l*=2 (quadrupole mode) are observed. However, there are controversies on the identification of modes [20-21]. A number of detailed studies were devoted to specific broadening and asymmetry of the phonon bands in the optical range of Raman spectra of $CdS_{1-x}Se_x$ nanoparticles [13] but lacks in the range of low frequency range particularly for the wide range of compositional intervals [22]. It is not clear whether the peaks appearing in the Raman spectra are *l*=0, *l*=2 and their overtones or torsional modes which should be absent according to the selection rule [23]. As far as, the overtones of the confined phonon modes (*n*>1) and the compositional dependent on low frequency phonon modes in the $CdS_{1-x}Se_x$ are concerned, to the best of our knowledge, there are only two reports [24-25]. However, they focus only on the overtones of symmetrical modes as they have been only able to observe them in their spectra. Moreover, they are not able to observe the torsional modes, which is quite obvious due to the selection rule for the perfect spherical nanoparticles [23]. But, their nanoparticles are not completely spherical [24-25].

In this paper, we present a systematic study of the low frequency Raman scattering from $CdS_{1-x}Se_x$ nanocrystals of various composition and sizes embedded in a borosilicate glass matrix. For this we have considered four samples having two different sizes of four different compositions. In order to identify the two additional peaks appearing in the experimental Raman spectra, the calculation of the low frequency phonon modes of $CdS_{1-x}Se_x$ nanoparticle embedded in borosilicate glass has been performed by using complex frequency model [25-28], core-shell model (CSM) [27,29,30] as well as classical Lamb model [19]. Out of these CFM and CSM consider the presence of medium, and estimates the damping of the vibrational modes resulting from the nanoparticle-medium interaction. These modes are used for distinguishing the overtone of spheroidal $l$=2, $n$=0 and torsional $l$=1, $n$=0 modes from the spheroidal $l$=0,$n$=0 and $l$=2,$n$=0 modes. The sizes obtained from the experimental spectra are in good agreement with these obtained from high-resolution transmission electron microscopy (HRTEM) [31]. The reason for the appearance of the forbidden torsional mode in the Raman spectra is discussed.

## II. EXPERIMENTAL DETAILS

$CdS_{1-x}Se_x$ nanocrystals were grown in a borosilicate glass matrix by a conventional technique of solid-state precipitation at temperatures of 600-800°C from a supersaturated solution formed by dissolving the initial reactants in the glass at high temperature and subsequent quenching [31]. Low frequency Raman spectra for all the four samples of different composition and sizes and are designated as CSS100, CSS82, CSS78 and CSS25 in the paper. The numbers next to the letters represent the concentration of selenium (Se) in the samples were recorded at room temperature in the backscattering geometry. A vertically polarized 488 nm line of an argon-ion laser (Coherent) with 500mW power was used to excite the samples. Unpolarized scattered light from the samples was dispersed using a double monochromator (Spex, model 14018) and detected using a photomultiplier tube (FW ITT 130) operated in the photon-counting mode. Scanning of the spectra and data acquisition was carried out using a microprocessor-based data-acquisition-cum-control system. For all the samples, low-frequency Raman spectra were recorded from 5 to 40 $cm^{-1}$ at 0.5 $cm^{-1}$ steps with 10 s integration time.

## III. THEORETICAL CONSIDERATIONS

Confined acoustic phonons in nanoparticles give rise to low frequency modes in the vibrational spectra of the materials. These low frequency phonon modes in borosilicate glass embedded $CdS_{1-x}Se_x$ nanoparticle can be obtained by solving the equation of motion of a homogeneous, free standing elastic sphere, first proposed by Lamb [19] and later used by several others in one or another forms [ 20-21, 24, 26-27].

However, to understand the key features of experimental Raman spectra and the vibrations of the $CdS_{1-x}Se_x$ nanoparticles are calculated by using three approach (i) using the original Lamb's approach [19] (ii) Solution of an embedded isotropic sphere by using complex frequency model [26-28] and (iii) core-shell model applicable to both free as well embedded nanoparticles [27,29-30]. Though the details of the original Lamb's approach can be found in existing literatures [19, 20], but for the sake of completeness and to make comparison with other two approaches, we describe briefly the model. The displacement field $\bar{u}(\bar{r},t)$ of an elastic medium of density $\rho(\bar{r})$ is governed by Navier's equation which is written as

$$c_{ijkl,j}u_{k,l} + c_{ijkl}u_{k,lj} = \rho\ddot{u}_i \qquad (1)$$

Where $c_{ijkl}(\bar{r})$ is the fourth rank elastic constant tensor field. However, the present equation for the homogenous, isotropic medium is written as [19]

$$(\lambda + 2\mu)\bar{\nabla}(\bar{\nabla}.\bar{u}) - \mu\bar{\nabla}\times(\bar{\nabla}\times\bar{u}) = \rho\ddot{u} \qquad (2)$$

Where, $\lambda$ and $\mu$ are Lame's constants and $\rho$ is the mass density of nanoparticle, which is related to each other by the expressions $V_l = \sqrt{(2\mu+\lambda)/\rho}$ and $V_t = \sqrt{\mu/\rho}$ , $V_l$ and $V_t$ are the longitudinal and transverse sound velocities in nanoparticles. Under stress free boundary conditions, Eq. (2) can be solved by introducing a scalar and vector potentials and yields two types of vibrational modes: spheroidal and torsional mode. These modes are described by orbital angular momentum quantum numbers $l$ and harmonic $n$. The eigen value equation for the spheroidal mode is expressed as [19]

$$2\left\{\eta^2+(l-1)(l+2)\left[\frac{\eta j_{l+1}(\eta)}{j_l(\eta)}-(l+1)\right]\right\}\frac{\zeta j_{l+1}(\zeta)}{j_l(\zeta)}-\frac{1}{2}\eta^4+(l-1)(2l+1)\eta^2+[\eta^2-2l(l-1)(l+2)]\frac{\eta j_{l+1}(\eta)}{j_l(\eta)}=0 \quad (3)$$

Here $\eta$ is the dimensionless eigenvalue expressed as

$$\eta_l^s=\frac{\omega_l^s R}{V_t} \quad (4)$$

And $\zeta$ is related to $\eta$ as $\zeta=\eta\left(V_t/V_l\right)$. The torsional mode is a vibration without dilatation, and its eigenvalue equation is given by [19]

$$j_{l+1}(\eta)-\left(l-1/\eta\right)j_l(\eta)=0 \quad \text{for } l=1 \quad (5)$$

The torsional modes are purely transverse in nature and independent of material property and are defined for $l$ =1 and are orthogonal to the spheroidal modes [19-20]. The spheroidal modes are characterized by $l$ =0, where $l$=0 is the symmetric breathing mode, $l$=1 is the dipolar mode and $l$=2 is the quadrupole mode. The $l$=0 mode is purely radial and produces polarized spectra, while $l$=2 mode is quadrupolar and produces partially unpolarized spectra. The spheroidal modes for even $l$ (i.e. $l$=0 and 2) are Raman active [23]. The lowest Eigen frequencies for $n$=0 for both spheroidal and torsional modes corresponds to the surface modes while for $n$ =1 corresponds to inner modes.

In the complex frequency model (CFM) [26-28], the nanoparticle is surrounded by a homogeneous and isotropic matrix of density $?_m$ and speeds of sound $V_{lm}$ and $V_{tm}$. The CFM is the result of fixed boundary condition and is different from the free boundary condition in the sense that it takes into account the stresses at the nanoparticles boundary, continuity of u and force balance at the nanoparticle-matrix interface due to the existence of the boundaries between particle and matrix interface. The boundaries between the particle and matrix modify the confined phonon modes and even some times responsible for the appearance of new modes. The boundary condition at large *R* is that u is an outgoing travelling wave. This model yields complex frequency, the real part *Re(? )* represents the frequency of the free oscillations. The imaginary part *Im(? ),* which is always positive, stands for the attenuation coefficient of the oscillations because the amplitude of the displacement is proportional to $e^{iRe(?)t-Im(?)t}$ at any place and therefore *Im(? )* yields damping. The real frequency, *Re(? )* in the present case are shifted from the frequency *?* obtained from Lamb equations. Therefore, the resulting complex nature of the frequency due to radiative decay enables to estimate both frequencies and their damping. Unlike the free vibrations of an elastic body the scalar and vector potentials include first kind of spherical Hankel functions, whose variable is purely imaginary in the region outside the nanoparticle in embedded systems. Verma et al [24] have also used a similar approach, to calculate the low frequency phonon modes of nanoparticle embedded in matrix. In experimental observations of such modes a broadened peak is observed with its centre at real frequency.

Though CFM model is able to predict the real frequency i.e. the actual peak position and broadening (FWHM) of the peak due to the damping actually observed in the Raman spectra but contains several inherent limitations. In these the most important one is the absence of valid wave functions required for the Raman scattering spectra calculations. In addition, the correspondence to quantum theory is also unclear due to the modes not being orthonormalizable and blowing up exponentially with the radial coordinate. To overcome this, Portalés et al [30] proposed an approach known as core-shell model (CSM) considering nanoparticles as core surrounded by a macroscopically large spherical matrix as shell which leads to real valued mode frequencies [28]. Therefore for calculating the absolute intensity or shape of low frequency Raman a spectrum, the CSM is the right approach for the comparison with the experimental Raman spectra. However, in a Raman scattering experiment the vibrational eigen frequencies for which the Raman intensity is related to the displacement inside the particle (core) of the systems are important [29].

In this model, the motion of nanoparticle, the mean square displacement $\langle u^2 \rangle_p$ is obtained as a function of mode frequency. This mean square displacement $\langle u^2 \rangle_p$ is nothing but the measure of the internal motion of the nanoparticle. The plots of $\langle u^2 \rangle_p$ with mode frequency are continuous and peaks in the plots correspond closely to the real part of the CFM. Half widths at half maximum of these peaks correspond closely to the imaginary parts of CFM frequencies [28]. Moreover, the Raman spectrum is governed by both $\langle u^2 \rangle_p$ and the electron-phonon interaction matrix element. The matrix element which appears as multiplicative constant, determines the overall intensity while $\langle u^2(\omega) \rangle_p$ determines both amplitude and the spectral line shape. Therefore, a preliminary step for obtaining the Raman spectrum would be to study the amplitude of nanoparticles vibrations as a function of phonon frequency. The mean square displacement inside the nanoparticles can be written as [29]

$$\langle u^2 \rangle_p = \frac{1}{v_p} \int_{R<R_p} \left\| \bar{u}(\bar{R}) \right\|^2 d^3 \bar{R} \tag{6}$$

Where, $v_p$ is the volume of the particle.

## IV. RESULTS AND DISCUSSION

Low-frequency Raman spectra from the samples CSS100, CSS82, CSS78 and CSS25 are presented in Fig.1. Raman scattering data are fitted to an exponential background and a Lorentzian line shape function. The results obtained from the fitting are presented in Table I along with the three different theoretical calculations such as Lamb's, CSM and CFM discussed in previous section. Since, the spectra are recorded in unpolarized configuration, both symmetrical (*l*=0) and quadrupolar (*l*=2) modes are expected to appear. All spectra show four peaks indicated by the indices 1,2, 3,and 4 can be seen at ˜ 10, ˜ 16, ˜ 21 and ˜ 27 $cm^{-1}$ respectively. These are attributed to the quadrupolar vibrations, torsional (dipolar) vibrations against the selection rule [23], overtone of quadrupolar vibrations and symmetrical vibrations in all four samples. The assignment of the modes is done with the help of theoretical calculations and on the comparison of the observed frequency ratio with the calculated. The calculated results on the low frequency phonon modes by using above discussed three approaches are depicted in table I along with the present Raman data. The table II presents the frequency ratio of the modes. As shown in Fig. 1 and Table I, the peak 1 at ˜ 10 $cm^{-1}$ and peak 4 at ˜ 27 $cm^{-1}$ are quadrupolar and symmetrical spheroidal mode vibrations respectively. The peaks 3 and 2 at ˜ 21 $cm^{-1}$ and at ˜ 16 $cm^{-1}$ respectively are overtone of *l*=2, spheroidal (*l*=2, 0 : SPH) i.e. *l*=2, *n*=1: SPH mode and torsional *l*=1, *n*=0 (*l*=1, 0 : TOR) mode. These all four peaks have been obtained for all four samples with variation in their frequencies due to 1/*R* size dependency. However for three samples CSS100, CSS78 and CSS25, the peak positions in the Raman spectra are almost same due to same size but different composition. These modes for sample CSS82 are at 11.34 $cm^{-1}$ and 5.5 $cm^{-1}$ respectively. While the overtone of *l* =2, 0 SPH is clearly seen the torsional mode *l* =1, 0 : TOR overlaps with the *l* =2, 0 : SPH mode in the case of sample CSS82. The overtone of the unpolarized mode (*l*=2,1 : SPH) has not been reported earlier. However, there are reports regarding the observation of overtones of the symmetrical mode (*l*=0, *n*=0 : SPH) [24-25]. While, Verma et al [24] pointed out the overlap of this mode with lower order symmetrical modes as the absence of *l*=2,1: SPH (overtone of quadrupolar mode). The Figure 2 presents the calculated frequency spectra by using CSM approach. Figure 1 and 2 along with the table I are clearly able to bring out the 1/*R* size dependency of the frequency. It is seen that the all approaches give more or less the same value of frequency for low frequency modes and compare well with the experimental values. The differences in the value of real frequency for modes obtained in three different approaches are due to the inclusion of effect of the medium in the case of CFM and CSM. The real frequency in the case of CFM corresponds to the peak position in CSM while the imaginary frequency is related to the broadening resulting from the interaction of nanoparticle and medium. The line width of the each mode observed in the Raman spectra along with the calculated line width for CSM and CFM are also listed in table I.

The most striking feature of the experimental Raman spectra is the appearance of *l*=2, *n*=1, SPH mode, which is overtone of *l*=2, *n*=0 : SPH mode and a torsional mode. In order to confirm this, we will take one by one, first considering the overtone of the quadrupolar mode appearing at ˜ 21 $cm^{-1}$ for CS100, CSS78 and CSS25 samples. This mode for the CSS82 sample is at ˜ 10 $cm^{-1}$. This mode has been assigned to the overtone of the quadrupole mode, *l* =2, *n*=0: SPH mode on the basis of the ratio $?_{20}/ ?_{21}$ presented in table II. The ratio of quadrupole mode to this mode $?_{20}/ ?_{21}$˜ 0.48 agree well with the ratio obtained for the same from the theoretical approaches. The CSM results presented in Figure 2 for both symmetrical (*l*=0) and quadrupole (*l*=2) modes have also been able to produce the overtone of *l*=2, *n*=0 mode similar to the experimental spectra for all three samples of equal size but with different compositions.

Another striking feature that is the observation of torsional (*l*=1, *n*=0 : TOR) mode at ˜ 16 $cm^{-1}$ for three samples of identical size of 2.2 nm. Although the torsional modes are forbidden in the Raman spectra, it could be due to the fact that the nanoparticles are not perfectly spherical [23,33]. To confirm this, we have calculated the frequency of *l*=1, *n*=0 : TOR mode and the ratio of *l*=2, *n*=0 : SPH to the *l*=1, *n*=0 : TOR mode, $?^{S}_{20}/ ?^{T}_{10}$, presented in table I and table II respectively, which agree reasonably well with the observed peak. Also, the appearance of torsional mode has been earlier confirmed by the Ovsyuk and Novikov [34] for nanoparticles embedded in $GeO_2$ glass, which they claim is due to the effect of surrounding matrix. However, we attribute the appearance of torsional mode to the non-spherical shape of nanoparticle as the presence of surrounding medium alone has not been able to produce the torsional mode for this mixed system [24, 25]. The non-spherical shape of the present glass embedded $CdS_{1-x}Se_x$ nanoparticle has been confirmed from high resolution transmission electron microscopy [31], In this case the selection rule is relaxed and the torsional vibrational mode with odd $l$ can be seen in the Raman spectra [23]. Absence of torsional modes $l$=1, $n$=0 : TOR for the sample CSSS82 in Raman spectra may be due to either overlap of this mode with $l$=2, $n$=0 : SPH mode or exact spherical shape of the nanoparticle. Former can be justified with fact that the calculated torsional mode $l$=1, $n$=0 : TOR frequency is close to the calculated $l$=2, $n$=0 : SPH mode.

Furthermore, the present results show that the frequency of quadrupolar mode ($l$=2, $n$=0 : SPH) remain more or less constant while the breathing mode ($l$=0, $n$=0 : SPH) shows some change with the composition.

## V. CONCLUSION

Low frequency acoustic phonon modes have been investigated for the $CdS_{1-x}Se_x$ nanoparticles embedded in the borosilicate glass using the low frequency Raman scattering and three different models namely classical Lamb's approach, the complex frequency model and core shell model. Based on the calculated spheroidal and torsional mode frequencies of nanoparticle the low frequency Raman spectra has been interpreted. A good agreement between theoretical and experimental results is obtained. The experimental Raman spectra show four peaks in the low frequency region. The torsional mode, forbidden by the selection rule for spherical particles has been observed and been attributed to the non-spherical shape of nanoparticles. In addition, the first harmonic of the quadrupolar mode has also been observed.

## ACKNOWLEDGEMENT

The financial assistance from BRNS, **Department of Atomic Energy** is highly appreciated.

## Figures Caption

**1.** Low frequency Raman spectra of $CdS_xSe_{1-x}$ nanoparticle embedded in borosilicate glass (a) CdSe (b) $CdS_{0.18}Se_{0.82}$ (c) $CdS_{0.22}Se_{0.78}$ and (d) $CdS_{0.25}Se_{0.75}$. Small circles, full curve and straight line are for the experimental Raman data, Curve obtained from fitting software peak fit and background respectively.

**2.** Low frequency spectra by using CSM for $CdS_xSe_{1-x}$ nanoparticle embedded in borosilicate glass (a) CdSe (b) $CdS_{0.18}Se_{0.82}$ (c) $CdS_{0.22}Se_{0.78}$ and (d) $CdS_{0.25}Se_{0.75}$.

## Tables Caption

1. Frequency for $CdS_xSe_{1-x}$ nanoparticle embedded in borosilicate glass from low frequency Raman spectra and theoretical models. In the embedded i.e. CSM and CFM case damping [Im (?)] is half width at half maxima. For Raman spectra the Im(?) is ? ?/2.
2. Comparison of frequency ratio. $?^S_{20}$ is the frequency of *l=2,n=0 ;* SPH mode and $?^T_{10}$ is the frequency of the *l=1,n=0;*TOR mode.

**Table I**

| SAMPLE No. | R (nm) | Type | Raman Shift (cm$^{-1}$) | | | | | | | |
|---|---|---|---|---|---|---|---|---|---|---|
| | | | Mode | Raman Expt. | | Lamb Model | CSM | | CFM | |
| | | | | Real ? | Im ? | | Real ? | Im ? | Real ? | Im ? |
| CSS100 | 2.2 | SPH | $?_{00}$ | 27.64 | 2.98 | 24.92 | 24.35 | 2.48 | 26.54 | 1.94 |
| | | | $?_{20}$ | 10.22 | 1.17 | 10.13 | 12.5 | 1.62 | 11.78 | 2.27 |
| | | | $?_{21}$ | 21.26 | 2.98 | 19.72 | 27.16 | 3.12 | 24.54 | 2.13 |
| | | TOR | $?_{10}$ | 15.66 | 2.91 | 14.20 | 17.60 | 3.89 | 17.20 | 3.11 |
| CSS82 | 4.65 | SPH | $?_{00}$ | 11.34 | 2.41 | 11.99 | 11.35 | 1.43 | 12.77 | 0.94 |
| | | | $?_{20}$ | 5.5 | 1.17 | 5.1 | 6.48 | 0.82 | 5.7 | 1.06 |
| | | | $?_{21}$ | 10.09 | 2.41 | 9.49 | 11.06 | 2.10 | 10.75 | 1.86 |
| | | TOR | $?_{10}$ | - | - | 6.24 | 6.9 | 1.94 | 6.72 | 2.06 |
| CSS78 | 2.2 | SPH | $?_{00}$ | 28.13 | 3.26 | 25.44 | 27.45 | 2.92 | 26.99 | 1.91 |
| | | | $?_{20}$ | 10.24 | 1.16 | 10.34 | 12.39 | 1.93 | 12.05 | 2.25 |
| | | | $?_{21}$ | 21.69 | 3.26 | 20.13 | 27.89 | 1.28 | 26.77 | 2.86 |
| | | TOR | $?_{10}$ | 15.66 | 2.91 | 14.20 | 17.60 | 3.89 | 17.20 | 3.11 |
| CSS25 | 2.2 | SPH | $?_{00}$ | 30.11 | 3.12 | 26.66 | 27.72 | 3.18 | 28.47 | 1.94 |
| | | | $?_{20}$ | 10.20 | 1.17 | 10.84 | 13.27 | 2.15 | 12.81 | 2.22 |
| | | | $?_{21}$ | 21.48 | 3.12 | 21.09 | 26.72 | 3.18 | 26.14 | 2.93 |
| | | TOR | $?_{10}$ | 15.66 | 2.91 | 14.20 | 17.60 | 3.89 | 17.20 | 3.11 |

**Table II**

| Composition $x$ | R (nm) | Samples | Expt. ratio | CSM ratio | CFM ratio | Lamb ratio |
|---|---|---|---|---|---|---|
| 1.0 | 2.2 | CSS100 | $?^{S}_{20}/?^{S}_{21}$=0.48<br>$?^{S}_{20}/?^{T}_{10}$=0.65 | $?^{S}_{20}/ ?^{S}_{21}$=0.46<br>$?^{S}_{20}/?^{T}_{10}$=0.71 | $?^{S}_{20}/?^{S}_{21}$=0.48<br>$?^{S}_{20}/?^{T}_{10}$=0.68 | $?^{S}_{20}/?^{S}_{21}$=0.51<br>$?^{S}_{20}/?^{T}_{10}$=0.71 |
| 0.82 | 4.65 | CSS82 | $?^{S}_{20}/?^{S}_{21}$=0.54<br>- | $?^{S}_{20}/?^{S}_{21}$=0.59<br>$?^{S}_{20}/?^{T}_{10}$=0.93 | $?^{S}_{20}/?^{S}_{21}$=0.53<br>$?^{S}_{20}/?^{T}_{10}$=0.84 | $?^{S}_{20}/?^{S}_{21}$=0.53<br>$?^{S}_{20}/?^{T}_{10}$=0.82 |
| 0.78 | 2.2 | CSS78 | $?^{S}_{20}/?^{S}_{21}$=0.47<br>$?^{S}_{20}/?^{T}_{10}$=0.64 | $?^{S}_{20}/?^{S}_{21}$=0.44<br>$?^{S}_{20}/?^{T}_{10}$=0.70 | $?^{S}_{20}/?^{S}_{21}$=0.45<br>$?^{S}_{20}/?^{T}_{10}$=0.70 | $?^{S}_{20}/?^{S}_{21}$=0.51<br>$?^{S}_{20}/?^{T}_{10}$=0.73 |
| 0.25 | 2.2 | CSS25 | $?^{S}_{20}/?^{S}_{21}$=0.48<br>$?^{S}_{20}/?^{T}_{10}$=0.64 | $?^{S}_{20}/?^{S}_{21}$=0.49<br>$?^{S}_{20}/?^{T}_{10}$=0.75 | $?^{S}_{20}/?^{S}_{21}$=0.49<br>$?^{S}_{20}/?^{T}_{10}$=0.74 | $?^{S}_{20}/?^{S}_{21}$=0.51<br>$?^{S}_{20}/?^{T}_{10}$=0.76 |

**Figure 1**

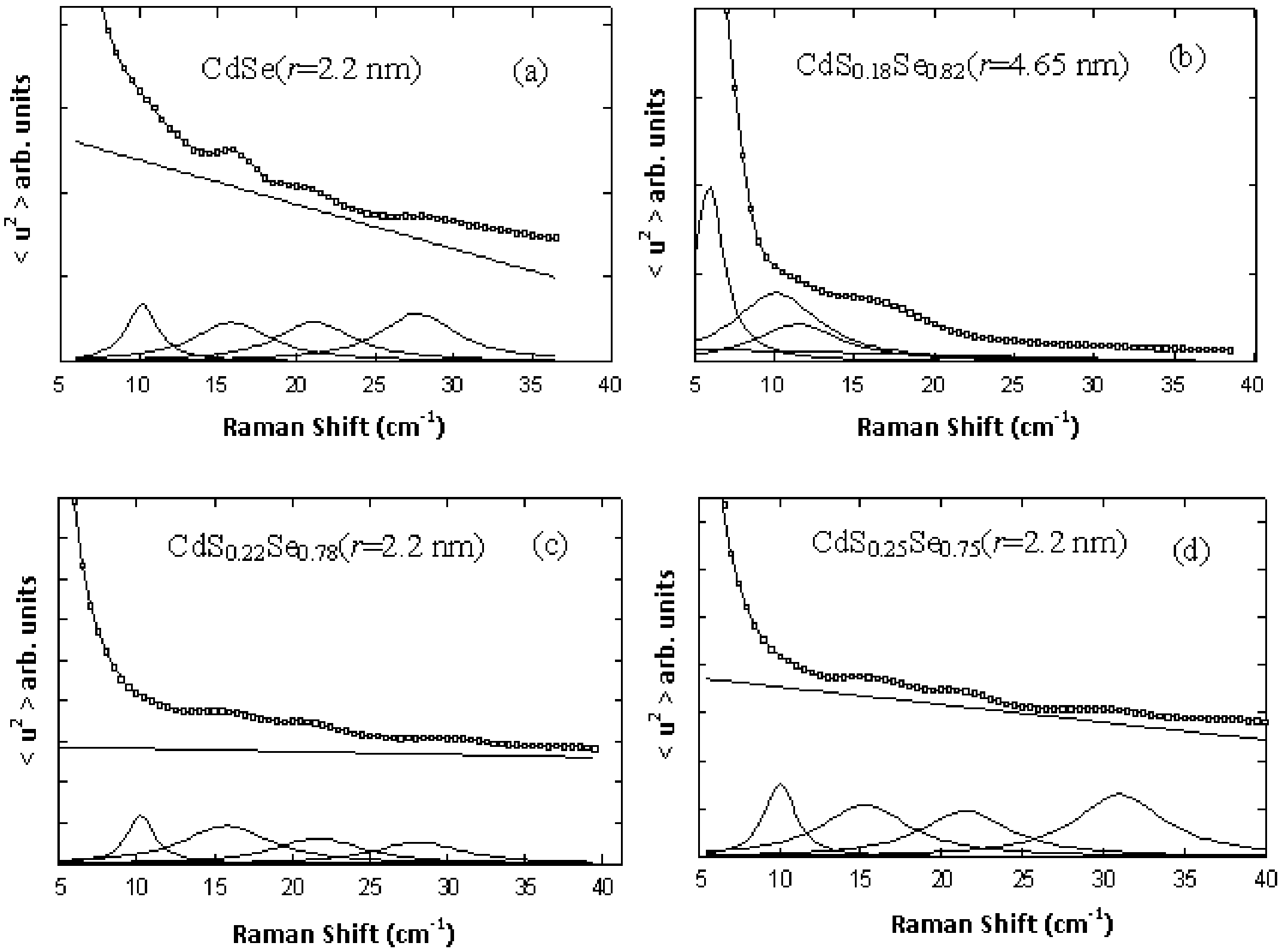

CdSe(r=2.2 nm)
(a)
< u2 > arb. units
Raman Shift (cm-1)
5 10 15 20 25 30 35 40
CdS0.18Se0.82(r=4.65 nm)
(b)
< u2 > arb. units
Raman Shift (cm-1)
5 10 15 20 25 30 35 40
CdS0.22Se0.78(r=2.2 nm)
(c)
< u2 > arb. units
Raman Shift (cm-1)
5 10 15 20 25 30 35 40
CdS0.25Se0.75(r=2.2 nm)
(d)
< u2 > arb. units
Raman Shift (cm-1)
5 10 15 20 25 30 35 40

**Figure 2**

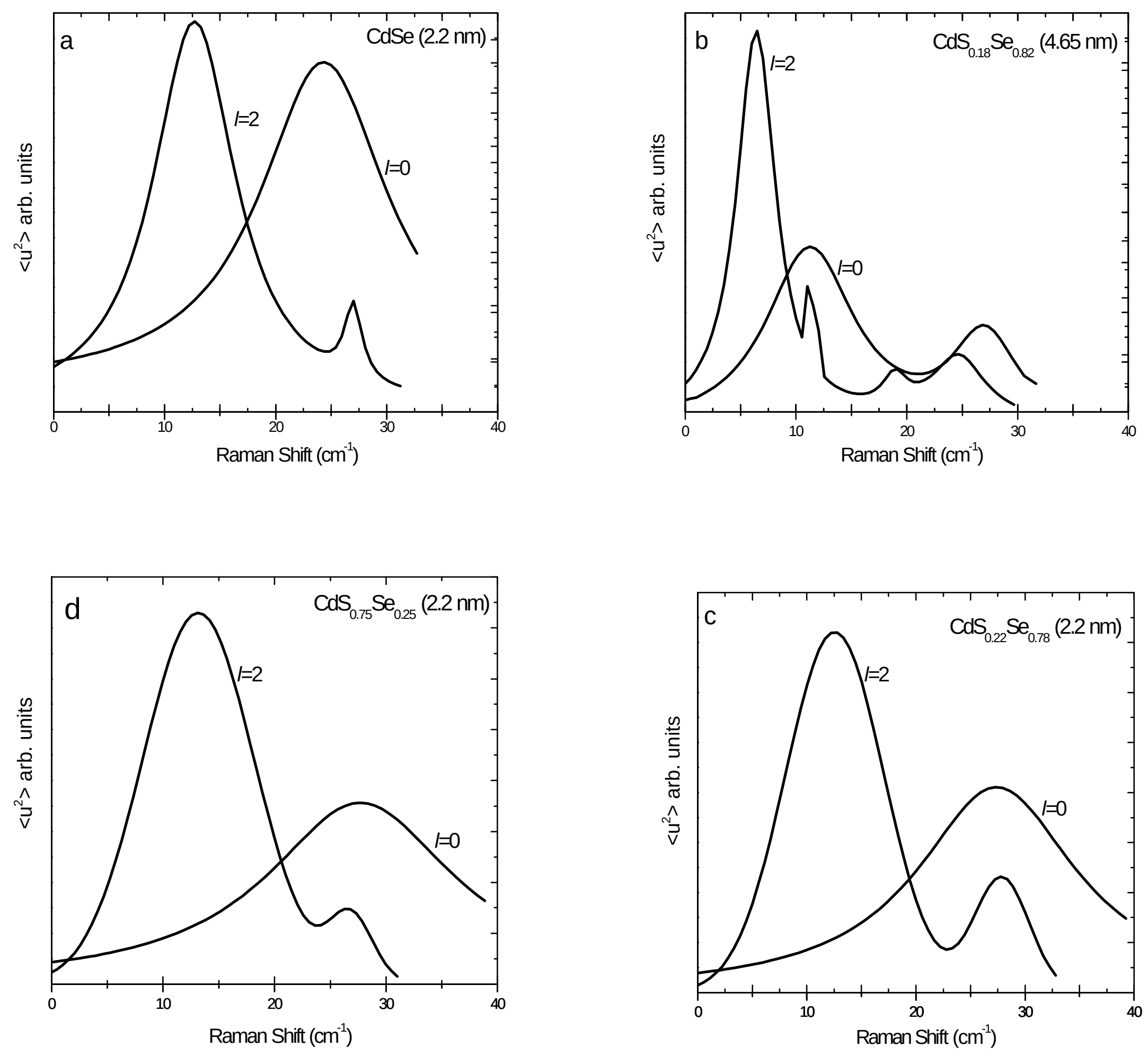